\documentclass[aps,pra,showpacs,twocolumn,superscriptaddress,a4paper,amsmath,amssymb,floatfix]{revtex4}

\usepackage{graphicx}
\usepackage{tabularx}
\usepackage{color}

\newcommand{\captionabove}[2][]{%
    \vskip-\abovecaptionskip
    \vskip+\belowcaptionskip
    \ifx\@nnil#1\@nnil
        \caption{#2}%
    \else
        \caption[#1]{#2}%
    \fi
    \vskip+\abovecaptionskip
    \vskip-\belowcaptionskip
}

\begin{document}

\title{Dissipative preparation of phase- and number-squeezed states with ultracold atoms}

\author{Roland Cristopher F. Caballar}
\affiliation{Asia Pacific Center for Theoretical Physics (APCTP), POSTECH,
San 31, Hyoja-dong, Nam-gu, Pohang, Gyeongbuk 790-784, Korea}
\affiliation{National Institute for Theoretical Physics (NITheP) and Centre for Quantum Technology, University of KwaZulu-Natal, Durban 4001, South Africa}
\affiliation{National Institute of Physics, University of the Philippines,
Diliman, Quezon City 1101, Philippines}
\author{Sebastian Diehl}
\affiliation{Institute for Theoretical Physics, University of Innsbruck,
A-6020 Innsbruck, Austria}
\author{Harri~M\"akel\"a}
\affiliation{QCD Labs, COMP Centre of Excellence, Department of Applied Physics, Aalto University,
P.O. Box 13500, FI-00076 AALTO, Finland}
\author{Markus Oberthaler}
\affiliation{Kirchhoff-Institute for Physics, University of Heidelberg, INF 227, 69120 Heidelberg, Germany}
\author{Gentaro Watanabe}\email{gentaro_watanabe@apctp.org}
\affiliation{Asia Pacific Center for Theoretical Physics (APCTP), POSTECH,
San 31, Hyoja-dong, Nam-gu, Pohang, Gyeongbuk 790-784, Korea}
\affiliation{Department of Physics, POSTECH, San 31, Hyoja-dong, Nam-gu, Pohang, Gyeongbuk 790-784, Korea}


\date{\today}

\begin{abstract}
We develop a dissipative quantum state preparation scheme for the creation of phase- and number-squeezed states. 
It utilizes  ultracold atoms in a double-well configuration immersed in a background Bose-Einstein condensate, with the latter consisting of an atom species different from the atoms in the double well and acting as a dissipative quantum reservoir. 
We derive a master equation for this system starting from microscopic physics, and show that squeezing develops on a time scale proportional to $1/N$, where $N$ is the number of particles in the double well. This scaling, caused by bosonic enhancement, allows us to make the time scale for the creation of squeezed states very short. The lifetime of squeezed states is limited by dephasing arising from the intrinsic structure of the setup. However, the dephasing can be avoided by stroboscopically switching the driving off and on. We show that this approach leads to robust stationary squeezed states. Finally, we provide the necessary ingredients for a potential experimental implementation by specifying a parameter regime for rubidium atoms that leads to squeezed states.
\end{abstract}


\maketitle

\section{Introduction}
The efficient generation of entanglement of macroscopic ensembles is a central challenge in atomic physics, both from a fundamental point of view and for practical applications such as quantum metrology and computation. One particularly prominent example of entanglement generation is provided by squeezed states. 
In  phase- and number-squeezed  states, the uncertainty in phase or particle number can be made as small 
as is compatible with the fundamental laws of quantum mechanics. In recent years, there has been considerable interest in squeezed states of ultracold atoms. This is motivated by their applications in matter-wave interferometry \cite{esteve, appel, cronin, leroux, gross, riedel, grond, lee}. 

A customary approach for the preparation of squeezed states involves unitary evolution and a measurement. An alternative strategy for quantum state preparation, based on the utilization of tailored dissipation, has been advocated recently \cite{poyatos, daley04, diehl, verstraete}.
In these schemes, a quantum bath is engineered and coupled to a system of interest in such a way that the system is driven into a desired target state. Advantages compared to more traditional state engineering techniques \cite{allothers} lie in the self-driven and deterministic character of the state preparation. Furthermore, the target state is reached starting from any initial state and the engineered dissipation can overwrite unwanted dissipation and decoherence mechanisms \cite{krauter}. 

Beyond the state preparation aspects, such a scheme opens up new scenarios for non-equilibrium many-body physics \cite{diehl2, weimer, diehlt} and quantum computation \cite{verstraete, pastawski, kastoryano}. Various platforms for the implementation of engineered dissipation have recently been explored theoretically \cite{kasto, reiter, chen, lukin, marcos, wstate, squeeze}.  Experimental realizations have been achieved in milestone experiments with atomic spin ensembles at room temperature \cite{krauter, muschikth, muschik} and systems of trapped ions \cite{barreiroexp, schindlerexp}. In Refs.~\cite{krauter, muschikth, muschik} it was shown that Einstein-Podolsky-Rosen-type entanglement of two distant atomic spin ensembles can be established dissipatively, while in Refs.~\cite{barreiroexp, schindlerexp}, Bell and Dicke states have been created deterministically.

In this work we propose a scheme for dissipative preparation of phase- and number-squeezed states using ultracold atoms.  It builds on a double-well geometry loaded with ultracold rubidium atoms (see Fig. \ref{fig_setup}), which is coherently driven in a double $\Lambda$-configuration, and immersed in a surrounding Bose-Einstein condensate (BEC)~\cite{note:atomspecies}. It realizes the Lindblad or quantum jump operators generating dissipative evolution proposed in Ref.~\cite{squeeze} for establishing macroscopic atomic entanglement. It extends the scheme proposed in Ref.~\cite{diehl} by an additional $\Lambda$-configuration, which is crucial for the implementation of the squeezing dynamics. Our scheme works if the number of particles in the double well, denoted by $N$, is macroscopic. We show that squeezing is established rapidly on a time scale $\tau_\gamma \sim 1/(N\gamma)$, where $\gamma$ is the overall rate in the master equation 
 extracted below  from a microscopic calculation, and $N$ is the number of particles in the double-well system. This rapid time scale originates from bosonic enhancement. 
In general, this setup leads to periodically oscillating squeezing.  We find that the oscillatory dynamics can be suppressed by switching the $\Lambda$-configuration 
lasers off and on periodically in time. Using this stroboscopic method, a robust and long-lived squeezed steady state can be created.

The rest of the paper is organized as follows. In Sec. \ref{sec:sys}, we briefly review the dissipative generation of atomic phase and number squeezing following Ref.~\cite{squeeze}, 
 explain the entangling mechanism, 
and outline a physical setup that realizes the dissipative dynamics leading to squeezed states. Based on a microscopic calculation, we derive and discuss the properties of the squeezing master equation in Sec. \ref{sec:master}, with details given in Appendices \ref{app:foh}-\ref{app:integk}.
In particular, by analyzing limiting cases analytically, we determine the time window  over which 
squeezed states exist. Analytical calculations yielding explicit expressions for the achievable amounts of squeezing within a mean-field approximation are corroborated and refined by numerical analysis in Sec. \ref{sec:results}.
In the long-time limit, an unavoidable dephasing effect will destroy the entanglement. 
We show  how this dephasing can be avoided using a stroboscopic method and demonstrate the conceptual feasibility of our scheme by presenting parameter values that lead to squeezed states in bosonic rubidium gases.  We present our conclusions in Sec. \ref{sec:conclusion}. We set $\hbar = 1$ throughout the paper.

\section{System for obtaining phase- and number-squeezed states \label{sec:sys}}

\subsection{Phase- and number-squeezing jump operators}
\label{sec:operators}

In Ref.~\cite{squeeze} we proposed an explicit form of squeezing jump operators. 
These operators are constructed such that their corresponding master equation has a phase- or number-squeezed state as a unique steady state.
Our goal in the present paper is to implement in a physical setup the squeezing jump operators to realize the dissipative preparation of phase- and number-squeezed states of cold atomic gases.

The squeezing jump operators were formulated for a two-state Bose system that has $N$ particles and is coupled to a suitable environment.
In this setting, assuming that the dynamics results entirely from the dissipation, such that the Hamiltonian of the system vanishes \cite{note:noham}, the dynamics of the density operator
$\hat{\rho}$ of the two-state system is determined by the master equation~\cite{squeeze}
\begin{equation}
\frac{d\hat{\rho}}{dt}=\frac{\gamma}{2}\left(2\hat{c}\hat{\rho}\hat{c}^\dagger-\left\{\hat{c}^\dagger\hat{c},\hat{\rho}\right\}\right),
\label{masteqsqz}
\end{equation}
where $\gamma$ is the dissipation rate, $\{\hat{A},\hat{B}\}\equiv\hat{A}\hat{B}+\hat{B}\hat{A}$, and $\hat{c}$ is a squeezing jump operator defined as
\begin{align}
\hat{c}&= (\hat{a}^\dagger_1+\hat{a}^\dagger_2)(\hat{a}_1-\hat{a}_2)+\nu(\hat{a}^\dagger_1-\hat{a}^\dagger_2)(\hat{a}_1+\hat{a}_2)\nonumber\\
&= 2(1+\nu)\hat{S}_z-2i(1-\nu)\hat{S}_y.
\label{sqzjumpop}
\end{align}
Here, $\hat{a}_i$ annihilates an atom in state $i=1,2$, $-1<\nu<1$ 
 is the parameter by which we can control the squeezing \cite{note:epsilon}, and we have introduced the SU(2) generators defined as $\hat{S}_x = (\hat{a}_1^\dagger\hat{a}_2+\hat{a}_2^\dagger\hat{a}_1)/2$, $\hat{S}_y = -i(\hat{a}_1^\dagger\hat{a}_2-\hat{a}_2^\dagger\hat{a}_1)/2$, and $\hat{S}_z=(\hat{a}_1^\dagger\hat{a}_1-\hat{a}_2^\dagger\hat{a}_2)/2$.

In Ref.~\cite{squeeze} we showed that, in the large-$N$ limit, Eq.~(\ref{masteqsqz}) evolves an arbitrary initial state towards a steady state that is phase or number squeezed provided the system is well described by the two-mode approximation.
If $\nu = 0$, the steady state of the master equation~\eqref{masteqsqz} is a coherent 
state defined as
\begin{align}
\psi_{\textrm{coh}}&=\frac{1}{\sqrt{2^NN!}}\, [\hat{a}_1^\dagger + \hat{a}_2^\dagger]^N |\mathrm{vac}\rangle \nonumber\\
&=2^{-\frac{N}{2}}\sum_{n=0}^{N} 
\sqrt{\frac{N!}{n! (N-n)!}}\, \left|n\right\rangle,
\end{align}
where $|\mathrm{vac}\rangle$ is the vacuum state and $|n\rangle$ is a Fock state with $n$ particles in the left well and $N-n$ particles in the right well.
For $\nu=0$, the coherent state is annihilated by $\hat{c}$, $\hat{c}\psi_{\textrm{coh}}=0$, and the system evolves towards $\psi_{\textrm{coh}}$ regardless of the initial state~\cite{diehl}. 
If $\nu\neq 0$, the steady state is a phase- or number-squeezed state. An explicit expression for this state can be found in Ref.~\cite{squeeze}.

The coherent state $\psi_{\textrm{coh}}$ is an eigenstate of $\hat{S}_x$ with the eigenvalue $N/2$, so $\langle\psi_{\textrm{coh}}|\hat{S}_x^2|\psi_{\textrm{coh}}\rangle=\langle\psi_{\textrm{coh}}|\hat{S}_x|\psi_{\textrm{coh}}\rangle^2=(N/2)^2$. For this reason, we call the approximation $\langle\hat{S}_x^2\rangle\equiv\textrm{Tr}[\hat{S}_x^2\hat{\rho}]\approx \langle\hat{S}_x\rangle^2$,  
where $\langle\hat{S}_x\rangle\approx N/2+O(N^0)$, the coherent-state approximation. 
It can be assumed to hold for the steady state corresponding to any value of $\nu$ provided $N\gg 1$ (see Ref.~\cite{squeeze} for discussion on the validity of this approximation).

The measures of phase and number squeezing are 
given, respectively, by
\begin{equation}
\xi_P\equiv\sqrt{\frac{2\langle\Delta\hat{S}^2_y \rangle}{|\langle\hat{S}_x \rangle|}}\ ,\;\;\;
\xi_N\equiv\sqrt{\frac{2\langle\Delta\hat{S}^2_z \rangle}{|\langle\hat{S}_x \rangle|}}\ ,
\label{evosqz}
\end{equation}
with $\langle\Delta\hat{S}^2_{y,z} \rangle = \langle\hat{S}^2_{y,z}\rangle-\langle\hat{S}_{y,z}\rangle^2$. To obtain the steady-state values for $\xi_P$ and $\xi_N$, we first calculate $d\langle \hat{S}_{y,z}^2\rangle/dt = \mathrm{Tr}[\hat{S}_{y,z}^2\, d\hat{\rho}/dt]$ ($d\langle \hat{S}_{y,z}\rangle/dt = \mathrm{Tr}[\hat{S}_{y,z}\, d\hat{\rho}/dt]$)
using Eq.~(\ref{masteqsqz}),
and then set the equation equal to zero to obtain the steady-state value for $\langle\hat{S}^2_{y,z}\rangle$ ($\langle\hat{S}_{y,z}\rangle$). In evaluating the equation, we use the approximations detailed in Ref.~\cite{squeeze}. These approximations will then be used to obtain an explicit form of $\langle\Delta\hat{S}^2_{y,z} \rangle$, which, together with the coherent-state approximation, is substituted in Eq.~(\ref{evosqz}). This gives the following expression for the steady-state value of $\xi_P$ and $\xi_N$:
\begin{equation}
\xi^{\rm SS}_{P,N}\simeq\sqrt{\frac{1\pm\nu}{1\mp\nu}}\ ,
\label{sssqueezingmeas}
\end{equation}
where the upper (lower) sign in both the numerator and denominator corresponds to the measure of phase (number) squeezing. 
A phase-squeezed steady state ($\xi^{\rm SS}_{P}<1$ and $\xi^{\rm SS}_{N}>1$) is obtained for $\nu<0$,  while a number-squeezed steady state ($\xi^{\rm SS}_{N}<1$ and $\xi^{\rm SS}_{P}>1$) is obtained for $\nu>0$. 

In the following we present a setup that yields the master equation (\ref{masteqsqz}) with the squeezing jump operator (\ref{sqzjumpop}) in the ideal limit. Then we show to what extent the physics of the squeezing jump operator can be realized in the actual setup.

\subsection{Description of the system}

\begin{figure}
\resizebox{8.2cm}{!}
{\includegraphics{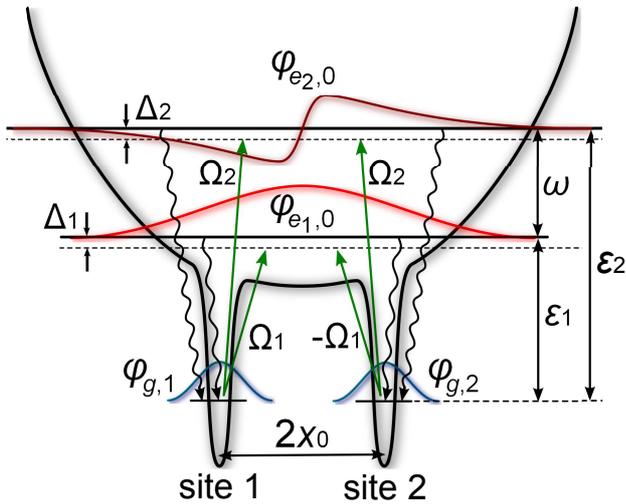}}
\caption{\label{fig_setup} (Color online) 
Schematic diagram of the setup for the realization of the squeezing
jump operator (\ref{sqzjumpop}). 
The two dips embedded in a wide harmonic potential form
a two-well configuration. Each of the wells holds one of the two degenerate 
ground states $\phi_{g,1}$ or $\phi_{g,2}$, which are identified
as the relevant states 1 and 2 for the squeezing jump operator, respectively.
The harmonic potential with frequency $\omega$ holds the 
even-parity state $\phi_{e_{1},0}$ and the odd-parity state $\phi_{e_{2},0}$.
The ground states $\phi_{g,1}$ and $\phi_{g,2}$ are coherently coupled to 
the excited state $\phi_{e_1,0}$ ($\phi_{e_2,0}$) using Raman lasers with 
antisymmetric (symmetric) Rabi frequencies $\Omega_1$ and $-\Omega_1$
($\Omega_2$ and $\Omega_2$), respectively (green solid arrows).
This driven setting is immersed in a background BEC,
which acts as a reservoir of Bogoliubov excitations, so that
atoms excited to the upper levels $\phi_{e_1,0}$ and $\phi_{e_2,0}$
can decay back to the ground states (black wavy arrows)
via spontaneous phonon creation in the reservoir.
In this setup the conditions 
$\sigma_g \alt x_0 \ll \sigma_e$ and $k_nx_0 \ll 1$ need to be satisfied. Here $\sigma_g$ and
$\sigma_e$ are the oscillator lengths of the well and the harmonic potential,
respectively, and $k_n$ is the wave number of the phonons created in the 
reservoir (see Sec.~\ref{sec:scales}). 
Efficient squeezing requires $\omega/(N\gamma) \ll 1$ as discussed
in detail in Sec.~\ref{sec:results}. 
It should be noted that this figure does not reflect the actual scales.
}
\end{figure}

In our setup we have $N$ bosonic atoms ($a$-atoms) with mass $m_a$ trapped in a quasi-one-dimensional (quasi-1D) external potential.  
This is a wide harmonic potential with two narrow wells embedded in it, as illustrated in Fig.~\ref{fig_setup}.  Our setup realizes a two-well configuration similar to the one used in the experiments described in Ref.~\cite{esteve}. 
The  wells have a characteristic frequency $\omega_{\rm well}$ and are located at $-x_0$ and $x_0$. Each well holds one of the two degenerate ground states $\phi_{g,1}$ and $\phi_{g,2}$ with the energy
$\epsilon_g\equiv \epsilon_{g,1} = \epsilon_{g,2}$. Here, the first subscript of $\phi$ and $\epsilon$ represents the energy level and the second subscript represents the site.  
These two states are identified as states 1 and 2, respectively, for the squeezing jump operator.
The harmonic potential 
has a characteristic frequency $\omega$ and energy levels $\epsilon_{e_{1},0}$ and $\epsilon_{e_{2},0}$ corresponding to an even-parity state $\phi_{e_{1},0}$ and an odd-parity state $\phi_{e_{2},0}$, respectively. 
Here the subscript ``0'' refers to the harmonic potential. The states in the two wells are populated
macroscopically with occupation number $\sim N$ (i.e., $\phi_{g,i}\sim \sqrt{N} \gg 1$ for $i=1,2$) while the states in the harmonic potential are not strongly occupied (i.e., $\phi_{e_{n},0}\sim 1$ for  $n=1,2$).

The ground states $\phi_{g,1}$ and $\phi_{g,2}$ of the wells are Raman coupled to the excited states $\phi_{e_{1},0}$ and $\phi_{e_{2},0}$
using lasers that are weak and far detuned.  The Raman lasers coupling $\phi_{g,1}$ and $\phi_{g,2}$ to $\phi_{e_{1},0}$
($\phi_{e_{2},0}$) have antisymmetric (symmetric) Rabi frequencies $\Omega_1$ and $-\Omega_1$ ($\Omega_2$ and $\Omega_2$), respectively,
and equal detuning $\Delta_1$ ($\Delta_2$).  This coherent coupling 
results in the annihilation part $\sim(\hat{a}_1 - \hat{a}_2)$
[$\sim(\hat{a}_1 + \hat{a}_2)$] in the squeezing jump operators (\ref{sqzjumpop}). 

This driven system is immersed in a BEC of a different species of bosonic atoms ($b$-atoms) with mass $m_b$, which works as a reservoir of Bogoliubov excitations \cite{diehl}. 
Atoms excited to $\phi_{e_1,0}$ and $\phi_{e_2,0}$ decay back to
$\phi_{g,1}$ and $\phi_{g,2}$ by spontaneously emitting Bogoliubov phonons into the background BEC.
These processes yield the creation parts of the squeezing jump operator (\ref{sqzjumpop}). 
The even (odd) parity of the state $\phi_{e_1,0}$ ($\phi_{e_2,0}$),
guaranteed by the symmetries of the harmonic-oscillator wave functions, 
leads to the creation part  $\sim (\hat{a}_1^\dag + \hat{a}_2^\dag)$
[$\sim (\hat{a}_1^\dag - \hat{a}_2^\dag)$] in Eq.~(\ref{sqzjumpop}). 
Here it is assumed that the decay events into $\phi_{g,1}$ and $\phi_{g,2}$ 
are indistinguishable (i.e., $k_n x_0\ll 1$ with $k_n$  being the wave number of emitted phonons;  this point will be discussed in detail in Sec.~\ref{sec:scales}).

The combination of the excitation and decay processes results in the creation of phase- and number-squeezed states. However, in order to maintain coherence between the two $\Lambda$-processes involving $\phi_{e_1, 0}$ and $\phi_{e_2, 0}$,  the spacing between $\epsilon_{e_1,0}$ and $\epsilon_{e_2,0}$ needs to be very small. This requires highly tunable and stable Raman beams, which can be realized with acousto-optical modulators.

There are two key differences in our setup compared to the scheme presented in Ref.~\cite{diehl}. 
First, there is an additional excited level $\phi_{e_2, 0}$. The absence of this level effectively corresponds to $\nu =0$, which reproduces the setting of the above reference. In this case, the steady state is the phase-locked symmetric superposition of the two lower wells, without any squeezing (i.e., a coherent state with relative phase 0).  Similarly, by switching off the coupling to the lower excited level $\phi_{e_1, 0}$ (formally achieved by sending $\nu \to \infty$) (see Ref.~\cite{note:epsilon}), a phase-locked antisymmetric superposition of the lower wells would be generated (i.e., a coherent state with relative phase $\pi$); the symmetry properties of the phase locking reflect the parity of the harmonic-oscillator wave functions as anticipated above.
However, only simultaneous coupling to the excited states allows for the preparation of squeezed states.   
The second difference is that here the occupation number $\sim N$ of each of the lower wells is macroscopic, while the setup in Ref.~\cite{diehl} targets optical lattices whose occupation number per site is typically of order one or less.
We will find theoretically in Sec. \ref{sec:master}, and validate numerically in Sec. \ref{sec:results}, that the large occupation of the quantum wells is important for the practical achievement of macroscopic squeezed states. In the long-time limit $t \to \infty$, the proposed setup will always suffer from dephasing which destroys the number or phase squeezing of the state due to the nonzero energy separation $\omega$ between the excited states. There will, however, be a time window  
$1/(N\gamma) < t < 1/\omega $ over which a squeezed state can be produced efficiently. Moreover, combining this setting with a stroboscopic element leads to robust stationary entangled states.

\subsection{Requirements on length and energy scales} 
\label{sec:scales}

In our setup we assume that the temperature of the background BEC is $T\simeq 0$. 
More precisely, the temperature is such that $T/\epsilon_n \ll 1$, where we have defined 
$\epsilon_n\equiv \epsilon_{e_n,0} -\epsilon_g$, $n=1,2$ (see Fig.~\ref{fig_setup}). 
By energy conservation, $\epsilon_n$ corresponds to the energy carried away by a phonon into the bath. As long as the relation $T/\epsilon_n \ll 1$ holds, the phonons will be emitted into an energy range where no excitations exist, so the BEC indeed acts as an effective zero-temperature reservoir~\cite{note:temperature}. We can therefore work with bath occupations $\langle\hat{b}_{\bf k}^\dagger \hat{b}_{\bf k}\rangle \simeq 0$ and $\langle\hat{b}_{\bf k} \hat{b}_{\bf k}^\dagger\rangle \simeq 1$, where $\hat{b}_{\bf k}$ annihilates a Bogoliubov excitation of momentum ${\bf k}$. 

Regarding the configuration of the trap, we assume that sites 1 and 2 hold only the well ground states $\phi_{g,1}$ and $\phi_{g,2}$, respectively (i.e., $\omega_{\rm well} \gg \epsilon_1$ and $\epsilon_2$). 
We also assume that the lasers coupling $\phi_{g,1}$ and $\phi_{g,2}$ to $\phi_{e_1,0}$ and $\phi_{e_2,0}$ are weak and far detuned, so $|\Omega_n/\Delta_n|\ll 1$, where $n=1$ and $2$. 
To eliminate any effect that higher-energy levels in the harmonic potential might have on the scheme, we require that $|\Delta_n/\omega|\ll 1$.

In addition, we require that $\omega_{\rm well}\gg \omega$ so that $\sigma_g \ll \sigma_e$, where $\sigma_g \equiv \sqrt{1/m_a\omega_{\rm well}}$ is the oscillator length of the well and $\sigma_e \equiv \sqrt{1/m_a\omega}$ is that of the harmonic trap. We further require that $\sigma_g \lesssim x_0$ so that the overlap between $\phi_{g,1}$ and $\phi_{g,2}$ is small and tunneling between the two wells can be neglected~\cite{note:overlap}.
Finally, we require a wide trap $x_0\ll\sigma_e$, which helps in making the trap energy levels $\epsilon_{e_1,0}$ and $\epsilon_{e_2,0}$ as close  as possible to each other.  
With these requirements, the three lengths are then ordered as follows: $\sigma_g\lesssim x_0\ll \sigma_e$.

To maintain interwell coherence in the spontaneous emission process, the condition $k_n x_0 \ll 1$ must be satisfied. Here $k_n$ is the
wave number of the phonons created in the reservoir. By energy conservation, it is determined by the condition $E_{k_n}=\epsilon_n\equiv\epsilon_{e_n,0}-\epsilon_g$,
where $E_k$ is the Bogoliubov excitation spectrum in the background BEC. The Bogoliubov spectrum is given by
\begin{equation}
E_{k} = \sqrt{\epsilon_{k}(2\rho_b U_0 + \epsilon_{k})}\, ,
\label{bog}
\end{equation}
where $\epsilon_{k} = k^2/2m_b$, $U_0 = 4\pi a_{bb}/m_b$, $a_{bb}$ is the intraspecies scattering length between two $b$-atoms, and $\rho_b$ is the number density of $b$-atoms. The condition $k_n x_0 \ll 1$ physically means that the wells 1 and 2 are indistinguishable from each other. 
Therefore, we obtain a coherent superposition of populations in these wells whose relative sign is determined by the symmetry of the harmonic-oscillator wave functions $\phi_{e,1}$ [a symmetric superposition $\sim (\hat{a}_1^\dag + \hat{a}_2^\dag)$ is created] and $\phi_{e,2}$ [an antisymmetric superposition $\sim (\hat{a}_1^\dag - \hat{a}_2^\dag)$ results]. 
The condition $k_n x_0 \ll 1$ can be satisfied by choosing $\epsilon_n$ comparable to $\omega$, i.e., $k_n \sim 1/\sigma_e$ (because $x_0 \ll \sigma_e$, we readily see that the condition $k_n x_0\ll 1$ holds if $k_n \sim 1/\sigma_e$).

As we consider a quasi-1D situation for simplicity, we need to assume that $\omega_\perp \gg \epsilon_n$ and $a_\perp \ll x_0$.  Here $\omega_\perp$ and $a_\perp\equiv \sqrt{1/m_a\omega_\perp}$ are the frequency and the oscillator length of the trap in the transverse directions, respectively.

\section{The Master Equation \label{sec:master}}
We start with the Hamiltonian for the trapped ultracold atomic gas, the background BEC, and the interaction between the trapped atomic gas and the BEC. The total Hamiltonian is given by
\begin{equation}
  \hat{H} = \hat{H}_a + \hat{H}_b + \hat{H}_{ab},
\end{equation}
with 
$\hat{H}_a=\sum_{n,i} \epsilon_{n,i} \hat{a}_{n,i}^\dagger \hat{a}_{n,i}$ the Hamiltonian of the trapped atoms ($a$-atoms), $\hat{H}_b = \sum_{\bf k}E_k \hat{b}_{\bf k}^\dagger \hat{b}_{\bf k}$
the Hamiltonian of the Bogoliubov excitations emitted into the BEC of $b$-atoms, and $\hat{H}_{ab}$ the Hamiltonian describing the interaction between the trapped atoms and the Bogoliubov excitations. Here $\hat{a}_{n,i}$ annihilates an $a$-atom in the state $\phi_{n,i}$. In the following, $n$ and $n'$ are the labels of the energy levels and $i$ and $i'$ are the labels of the sites. 
The possible values of the indices are $n,n'= g,e_1,e_2$ and $i,i'=0,1,2$.

The term $\hat{H}_{ab}$ is obtained using the field operators for the trapped atoms and the background BEC, with their explicit forms given, together with further details on the derivation of $\hat{H}_{ab}$, in Appendix \ref{app:foh}. Explicitly, $\hat{H}_{ab}$ has the form
\begin{equation}
  \hat{H}_{ab} 
\simeq \sum_{{\bf k} \ne 0} g_k \hat{A}_{\bf k}^\dagger \hat{b}_{\bf k} 
+ \mbox{h.c.}\ ,
\label{inthamil}
\end{equation}
with
\begin{align}
  g_k \equiv& \frac{2\pi a_{ab}}{\mu} \sqrt{\rho_b} S_k^{1/2},\quad
  S_k \equiv \frac{k^2}{2m_b E_k},\label{gkn}\\
  \hat{A}_{\bf k}^\dagger \equiv&\ e^{-k_\perp^2 a_\perp^2 /4}
\sum_{\substack{n,n'\\i,i'}} {\cal A}_{k_x; i,i'}^{(n,n')}\ \hat{a}_{n,i}^\dagger \hat{a}_{n',i'}.\label{skn}
\end{align}
Here $a_{ab}$ is the interspecies scattering length between the $a$-atoms and the $b$-atoms, $k_\perp$ and $k_x$ are the wave numbers in the transverse directions and the $x$ direction respectively, $a_\perp$ is the oscillator length of the trap in the transverse directions, and $\mu=(m_a + m_b)/m_a m_b$ is the reduced mass.
The operator $\hat{A}_{\bf k}^\dagger$ describes the transition of an $a$-atom from $\epsilon_{n',i'}$ to $\epsilon_{n,i}$ due to an interaction between an $a$-atom and a Bogoliubov excitation~\cite{diehl}. The functions ${\cal A}_{k_x; i,i'}^{(n,n')} \equiv \int dx\ e^{ik_x x} \phi_{n,i}^*(x)\phi_{n',i'}(x)$, appearing in Eq.~(\ref{skn}), are the overlap integrals between the states $\phi_{n,i}(x)$ and $\phi_{n',i'}(x)$. Details about their evaluation are provided in Appendix \ref{app:overlapint}.

Using the Born-Markov approximation, the second-order master equation can be written as
\begin{equation}
  \dot{\hat{\rho}}(t) = {\cal L}\hat{\rho} = -\int_0^\infty dt'\ 
\mathrm{Tr}_R\left[\hat{H}_{ab}(t),[\hat{H}_{ab}(t-t'),
\hat{\rho}(t)\otimes\hat{R}]\right],
\end{equation}
where $\mathrm{Tr}_{R}$ is the trace over the background BEC variables and $\hat{R}$ the density matrix of the reservoir. Following  Ref.~\cite{daley04}, we obtain the master equation
\begin{align}
  \dot{\hat{\rho}}(t) \simeq & \sum_{\bf k}\pi |g_{\bf k}|^2 \delta(E_{\bf k}-\epsilon_n)\left\{
[\hat{A}_{\bf k}(t),\hat{\rho}\hat{A}_{\bf k}^\dagger(t)] + \mbox{h.c.}
\right\},
\label{master1}
\end{align}
where
\begin{align}
  \hat{A}_{\bf k}^\dagger(t) =& e^{-k_\perp^2 a_\perp^2 /4}\sum_{n,i=1,2}
e^{i\epsilon_n t}{\cal A}_{k_x;0,i}^{(e_n,g)}\hat{a}_{e_n,0}^\dagger \hat{a}_{g,i}.
\label{interterm1}
\end{align}
In deriving the master equation, we use the assumption that $T\approx 0$ for the background BEC 
so that the trapped $a$-atoms are not excited to higher-energy states by $\hat{H}_{ab}$. 
Also, since the occupation number of the lowest-energy well states $\phi_{g,i}$ is of order $N$ and that of the harmonic-trap energy states $\phi_{e_1,0}$ and $\phi_{e_2,0}$ is of order unity, terms corresponding to the transition $e_n\rightarrow g$ are a factor $\sqrt{N}$ larger than those  
corresponding to the transition $e_2\rightarrow e_1$. 
Consequently, assuming that $N\gg 1$, the dominant terms in $\hat{A}_{\bf k}^\dagger$ are $\hat{a}_{e_1,0}^\dagger \hat{a}_{g,i}$ and $\hat{a}_{e_2,0}^\dagger a_{g,i}$ and their conjugates.  These are the terms appearing in the master equation~\eqref{master1}.

We replace the summation over ${\bf k}$ with an integration over ${\bf k}$ in Eq.~(\ref{master1}). After integrating, we adiabatically eliminate the harmonic trap energy states $\hat{a}_{e_n,0}$, with details for the integration and adiabatic elimination given in Appendix \ref{app:integk}. Finally, we obtain the following form for the master equation: 
\begin{widetext}
\begin{align}
{\cal L}\hat{\rho} \simeq& \frac{\gamma}{2}\Big\{
\left(2\hat{c}^\dagger_+ \hat{c}_-\hat{\rho}\hat{c}_-^\dagger \hat{c}_+ - \left\{\hat{c}^\dagger_- \hat{c}_+\hat{c}^\dagger_+ \hat{c}_-,\hat{\rho}\right\}\right)
  +\eta\nu^2\left(2\hat{c}^\dagger_- \hat{c}_+\hat{\rho}\hat{c}^\dagger_+ \hat{c}_- - \left\{\hat{c}^\dagger_+ \hat{c}_-\hat{c}^\dagger_- \hat{c}_+,\hat{\rho}\right\}\right)\nonumber\\
&\quad  +\nu\left(e^{-it(\epsilon_2 - \epsilon_1)}[\hat{c}^\dagger_- \hat{c}_+,\hat{\rho}\hat{c}^\dagger_- \hat{c}_+]\right.
  +\left.e^{it(\epsilon_2 - \epsilon_1)}[\hat{c}^\dagger_+ \hat{c}_-\hat{\rho},\hat{c}^\dagger_+ \hat{c}_-]\right)\nonumber\\
&\quad  +\eta\nu\left(e^{-it(\epsilon_2 - \epsilon_1)}[\hat{c}^\dagger_- \hat{c}_+\hat{\rho},\hat{c}^\dagger_- \hat{c}_+]\right.
  +\left.e^{it(\epsilon_2 - \epsilon_1)}[\hat{c}^\dagger_+ \hat{c}_-,\hat{\rho}\hat{c}^\dagger_+ \hat{c}_-]\right)\Big\},
\label{masteq0}
\end{align}
\end{widetext}
where
\begin{align}
\eta(k_1,k_2)\equiv& k_2\gamma_2/k_1\gamma_1,\label{etaconst}\\
  \gamma \equiv& 4\sqrt{\pi}\,k_1\sigma_g\gamma_1\phi_{e_1,0}^2(x_0)
\left(\frac{\Omega_1}{\Delta_1}\right)^2,\label{gamconst}\\
\gamma_n\equiv& \frac{1}{(2\pi)^3}\frac{2\pi^2 k_n |g_{k_n}|^2}{v_n},\quad v_n \equiv \left.\frac{\partial E_{\bf k}}{\partial k}\right|_{k=k_n},\label{gamman}\\
\nu \equiv& \frac{\Omega_2}{\Omega_1}\frac{\Delta_1}{\Delta_2}\,
\tilde{\nu},\quad \tilde{\nu} \equiv\frac{\phi_{e_2,0}(x_0)}{\phi_{e_1,0}(x_0)},\label{epsi}\\
\hat{c}_+ \equiv& \hat{a}_{g,1}+\hat{a}_{g,2},\quad \hat{c}_- \equiv \hat{a}_{g,1}-\hat{a}_{g,2}.\label{cop}
\end{align}
We learn from these relations that any value of the squeezing parameter $\nu$ can be generated in our setup by adjusting the relative strengths of the Rabi frequencies and detunings. Note, however, that this has to be done in a fashion that does not invalidate the scale hierarchies described 
in Sec.~\ref{sec:scales}. 
If $\epsilon_{n}/\rho_b U_0\ll 1$ for $n=1,2$ 
as in the case that we will
consider in Sec.~\ref{sec:implement}, $\eta$ is well approximated as
\begin{equation}
  \eta(k_1,k_2) \simeq \left(\frac{\epsilon_2}{\epsilon_1}\right)^3=\left(1+\frac{\omega}{\epsilon_1}\right)^3.
\end{equation}
Comparing Eq.~(\ref{masteq0}) to the master equation in Ref.~\cite{diehl}, we see that there are two fundamental differences. The first is the presence of $\nu$ in Eq.~(\ref{masteq0}), which is due to the use of two different $\Lambda$-processes in our scheme.  
The second is the presence of the time-dependent exponential factors $e^{\pm it(\epsilon_2 - \epsilon_1)}$ following from the nonzero separation $\omega\equiv\epsilon_2-\epsilon_1 $ between the first two energy levels of the harmonic trap.

To bring out the physics of this equation more clearly, let $C_{11}=1$, $C_{22}=\eta\nu^2$, $C_{12}=\nu$, and $C_{21}=\eta\nu$. Also, let
\begin{equation}
\hat{F}_1\equiv \hat{F}_2^\dagger\equiv \hat{c}^\dagger_+ \hat{c}_-.
\end{equation}
We can then write Eq.~(\ref{masteq0}) as 
\begin{equation}
{\cal L}\hat{\rho} \simeq \frac{\gamma}{2}\sum_{n,n'=1,2}C_{nn'}\left(e^{-it(\epsilon_n - \epsilon_{n'})}[\hat{F_n}\hat{\rho},\hat{F}_{n'}^\dagger]+\mathrm{h.c.}\right).
\label{masteq1}
\end{equation}
This makes it clear that there is no frame of reference where the energies could be gauged away in general. Let us now study important limiting cases of this equation. We first consider the limit  $\epsilon_2 - \epsilon_1 \to 0$, which implies  $k_2\rightarrow k_1$ and $\gamma_2\rightarrow\gamma_1$, so that in turn $\eta\rightarrow 1$. Then the explicit time dependence vanishes and the matrix $C$ takes a factorized form $C = \vec v \vec v^T$, with $\vec v = (1,\nu)^T$. In this, and only this case, the jump operators do not appear in an incoherent sum of both processes $F_1$ and $F_2$, but they rather take the form of a coherent superposition
\begin{align}
  \hat{c} \equiv& \vec v^T\cdot \vec F =  \hat{F}_1 + \nu \hat{F}_2\nonumber\\
=& (\hat{a}^\dagger_1 + \hat{a}^\dagger_2)(\hat{a}_1-\hat{a}_2)
+ \nu (\hat{a}^\dagger_1 - \hat{a}^\dagger_2)(\hat{a}_1+\hat{a}_2).
\label{squeezejumpop}
\end{align} 
The master equation 
(\ref{masteq1}) then reduces to the form anticipated in Sec.~\ref{sec:operators}:
\begin{equation}
  {\cal L}\hat{\rho} \simeq \frac{\gamma}{2}
\left(2\hat{c}\hat{\rho}\hat{c}^\dagger - \hat{c}^\dagger\hat{c}\hat{\rho}
- \hat{\rho}\hat{c}^\dagger\hat{c}\right),
\label{squeezemasteq}
\end{equation}
where $\hat{c}$ is the squeezing jump operator given in Ref.\ \cite{squeeze}.

In contrast, if $\epsilon_2-\epsilon_1\rightarrow\infty$, we can neglect the last two terms in Eq.\ (\ref{masteq0}) due to the rapidly oscillating exponential factors $e^{\pm it(\epsilon_2-\epsilon_1)}$ so that the master equation will have the following form:
\begin{align}
{\cal L}\hat{\rho} \simeq& \frac{\gamma}{2}\Big[
\left(2\hat{F}_1 \hat{\rho}\hat{F}^\dagger_1 - \left\{\hat{F}^\dagger_1 \hat{F}_1,\hat{\rho}\right\}\right)\nonumber\\
& +\eta\nu^2\left(2\hat{F}_2 \hat{\rho}\hat{F}^\dagger_2 - \left\{\hat{F}^\dagger_2 \hat{F}_2,\hat{\rho}\right\}\right)\Big].
\label{masteq2}
\end{align}
This master equation describes two incoherent processes, each of them similar to those described in Ref.\ \cite{diehl}. The processes described by the first and second lines of Eq.~(\ref{masteq2}) are used to prepare phase-locked states with relative phase of $0$ and $\pi$, respectively.

This discussion makes it clear that at long times $t(\epsilon_2-\epsilon_1) \gg 1$, 
the energy difference between the excited levels will unavoidably lead  
to dephasing of the squeezed state. It is therefore crucial to compare the time scale 
\begin{align}
T_\omega =\frac{2\pi}{\omega}=\frac{2\pi}{\epsilon_2 - \epsilon_1}
\end{align}
 of the trapping potential, which marks the onset of dephasing, to the time scale $\tau_\gamma$ over which squeezing builds up. The latter is obtained from the equation of motion for $\langle\hat{S}^2_{y,z}\rangle$ discussed in Ref.~\cite{squeeze} and is given by
\begin{equation}
\tau_{\gamma} =\frac{1}{4N\gamma(1-\nu^2)}.
\label{chartime}
\end{equation} 
Crucially, the \emph{effective rate for squeezing} $ \gamma_{\text{eff}}= \tau_\gamma^{-1}  = 4N\gamma(1-\nu^2) \propto N$ is proportional to the number of atoms trapped in the wells \cite{note:epsilon1}. 
This can be traced back to bosonic amplification.  A squeezed state is then generated in the time window 
\begin{align}
\label{time_window}
\tau_\gamma <  t < T_\omega,
\end{align}
in which nontrivial quantum mechanical correlations have built up, but dephasing is still negligible. 

Nonzero energy difference $\omega$ between the excited levels is unavoidable in our setting if the scale hierarchy of Sec.~\ref{sec:scales} is to be respected. 
This has two physically distinct effects: First, it introduces dephasing as explained above, and second, it leads to $\eta\neq 1$ [see Eq.~\eqref{etaconst}]. 
While dephasing will always destroy entanglement in the long-time limit, a small deviation of $\eta$ from unity will still allow for phase- and number-squeezed states [see Eq.~\eqref{sssqueezingmeaszeroom} below]. 
This motivates us to consider a modified continuous evolution with a stroboscopic element, where the evolution is  interrupted at intervals shorter than $T_\omega$ and is immediately restarted. In this way, we disentangle the two effects of the finite excited level spacing. It will allow us to generate effectively stationary entanglement between the two wells. 

\section{Numerical Results \label{sec:results}}
\subsection{Properties of the master equation}
We now study the dynamics numerically.  
An important quantity characterizing the density operator is the purity $P[\hat{\rho}]$, 
\begin{equation}
P[\hat{\rho}]=\frac{1}{N}\left\{(N+1)\mathrm{Tr}\left[\hat{\rho}^2\right]-1\right\}.
\label{puri}
\end{equation}
For a pure state $P[\hat{\rho}]=1$ and for the maximally mixed state $P[\hat{\rho}]=0$. It is in general preferable to maximize the purity of squeezed states. As can be seen from Eq.~\eqref{time_window}, the dimensionless ratio $\tau_\gamma/T_\omega\propto \omega/N\gamma$ characterizes the time interval during which the system is in a squeezed state. 
For all the numerical calculations shown below, we take a coherent state with relative phase 0 as the initial state. 
The main conclusions do not change if another initial state is used.

\begin{figure}
\includegraphics[scale=0.85]{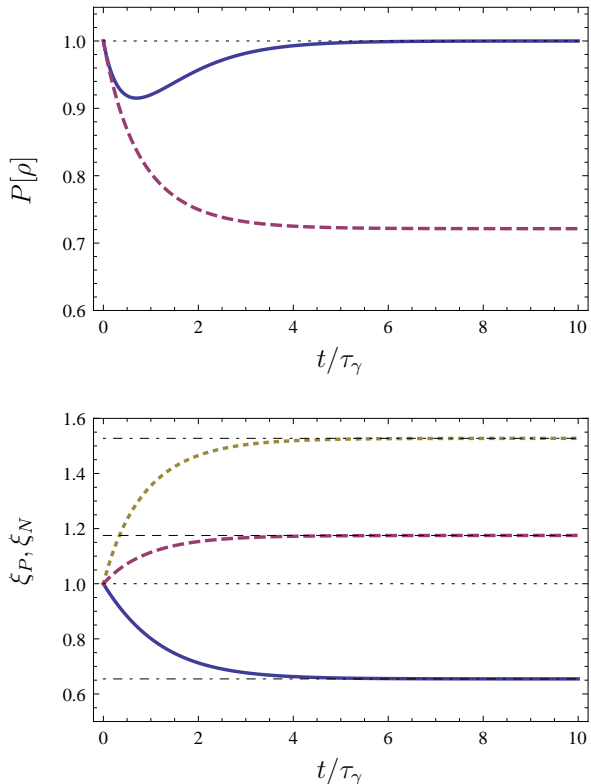}
\caption{ 
(Color online) 
Purity and measures of phase and number squeezing for the ideal case corresponding to Eq.~\eqref{squeezemasteq} and the completely incoherent case corresponding to Eq.~\eqref{masteq2}. 
In the top panel, the solid (dashed) line refers to the ideal (incoherent) case. 
In the bottom panel, the solid (dotted) line gives $\xi_P$ ($\xi_N$) for the ideal case. 
For the incoherent case, $\xi_P$ and $\xi_N$ are equal and are denoted by the dashed line. 
The horizontal dash-dotted and dashed lines show the theoretical steady-state values of 
squeezing given by Eqs.~\eqref{sssqueezingmeas} and \eqref{inco}, respectively. 
Here $N=100$, $\gamma>0$ is arbitrary, $\nu=-0.4$, and $\eta=1$. 
\label{fig:ideal_and_incoherent}
}
\end{figure}
\begin{figure}
\includegraphics[scale=0.85]{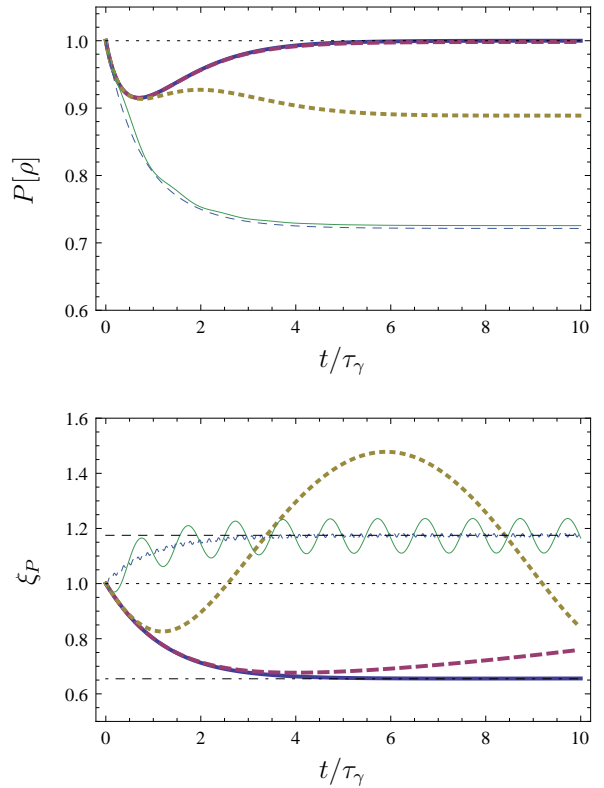}
\vspace{5mm}
\caption{(Color online) Purity and measure of phase squeezing for a system with 
 $N=100$, $\nu=-0.4$, and $\eta=1$. The solid, dashed, dotted, thin solid, and thin dashed lines correspond to $\tau_\gamma/T_\omega=0.001$, $0.01$, $0.1$, $1$, and $10$, respectively. 
 The horizontal dash-dotted and dashed lines show the theoretical steady-state values of 
squeezing corresponding to the ideal case [Eq.~\eqref{sssqueezingmeas}] and completely incoherent case [Eq.~\eqref{inco}], respectively. The larger $\tau_\gamma/T_\omega$ is, the earlier $\xi_P$ starts to deviate from the behavior of the ideal case. 
}
\label{fig:transition_to_incoherent}
\end{figure}

We study first the behavior of the purity and the measure of phase squeezing in the  
ideal case, $\tau_\gamma/T_\omega=0$ and $\eta=1$ [Eq.~\eqref{squeezemasteq}], 
and in the completely incoherent case, $\tau_\gamma/T_\omega=\infty$  [Eq.~\eqref{masteq2}]. 
Here and in what follows the term ``ideal'' refers to the simplified master equation \eqref{squeezemasteq} studied in Ref.~\cite{squeeze}.
The time evolution is shown in Fig.~\ref{fig:ideal_and_incoherent}.  
In the ideal case, $\hat{\rho}$ evolves towards a pure phase- or number-squeezed state. 
The steady-state values of $\xi_{P,N}$, indicated by the horizontal dash-dotted lines, are given by Eq.~\eqref{sssqueezingmeas}. In the case of the incoherent master equation \eqref{masteq2}, the steady state is a mixed state. The measures $\xi_P$ and $\xi_N$ are equal throughout the time evolution and approach the steady-state value 
given by  
\begin{equation}
\xi_{P}^{{\rm SS},\rm{inco}}=\xi_{N}^{{\rm SS},\rm{inco}}=\sqrt{\frac{1+\eta\nu^2}{1-\eta\nu^2}}.
\label{inco}
\end{equation}
This expression is derived in a manner similar to that used to derive Eq.~(\ref{sssqueezingmeas}), except that $d\hat{\rho}/dt$ is calculated using Eq.~(\ref{masteq2}) instead of Eq.~(\ref{masteqsqz}). Note that $\xi_{P}^{{\rm SS},\rm{inco}}$ and $\xi_{N}^{{\rm SS},\rm{inco}}$ are both greater than or equal to $1$.

In a realistic experimental setup, $\tau_\gamma/T_\omega$ is finite and the 
dynamics lies somewhere between the two extreme cases studied above. 
We illustrate the dependence of the purity and squeezing on the ratio $\tau_\gamma/T_\omega$  
in Fig.~\ref{fig:transition_to_incoherent}. 
By comparing the solid lines in Figs.~\ref{fig:ideal_and_incoherent} and \ref{fig:transition_to_incoherent}, 
we see that at $\tau_\gamma/T_\omega = 0.001$,  the time evolution of $P$ and $\xi_P$ is very close to  that of the ideal system during the time interval shown in the figure. 
As $\tau_\gamma/T_\omega$  increases, the dynamics becomes significantly different from the ideal case, as shown by the dashed, dotted, and thin solid curves in Fig.~\ref{fig:transition_to_incoherent}. 
The  purity decreases and $\xi_{P,N}$ oscillates with an oscillation period $T_\omega$. The oscillatory behavior can be attributed to the time-dependent phase factors $e^{\pm i\omega t}$ appearing in Eq.~\eqref{masteq1}. These factors will cause $\hat{\rho}$ to dephase, leading to the loss of purity.  An approach to suppress this dephasing will be detailed in the next section. Despite the oscillations, squeezing can be maintained for an interval of time equal to half of the oscillation period. Consequently, even under nonideal conditions $(\omega>0)$, it is possible to observe squeezing over experimentally relevant time scales if $T_\omega$ is sufficiently large and $\tau_\gamma$ is sufficiently small. This is promising for the experimental implementation of our state preparation scheme. Around $\tau_\gamma/T_\omega\approx 1$, the behavior of $P$ and $\xi_P$ begins to resemble that described by the incoherent master equation [Eq.~\eqref{masteq2}].

\begin{figure}
\includegraphics[scale=0.9]{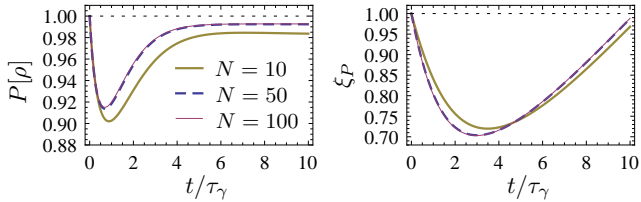}
\vspace{1mm}
\caption{(Color online) Purity and measure of phase squeezing for different values of the particle number.  
Here $\tau_\gamma/T_\omega=0.02, \nu=-0.4$, and $\eta=1$. 
}
\label{fig:N_dependence}
\end{figure}

Finally, we briefly discuss the $N$ dependence of the purity and the measure of squeezing for fixed $\tau_\gamma/T_\omega$. 
In experiments, $N$ can be on the order of $10^3$ or larger. Numerical calculations 
with this large particle number are very time consuming. 
Fortunately, the particle number dependence 
of squeezing and purity becomes weak already at $N\approx 100$. 
That is, the dynamics of these quantities for a fixed $\tau_\gamma/T_\omega$ is almost independent of $N$ if $N$ is comparable to or larger than $100$. 
We illustrate this  in Fig.~\ref{fig:N_dependence}, where the measure of phase squeezing and purity corresponding to $N=50$ and $100$ can be seen to be nearly identical. 
In all the numerical simulations reported in this article, we have chosen $N=100$.

\subsection{Implementation with ultracold atoms\label{sec:implement}} 
\begin{figure}
\includegraphics[scale=0.90]{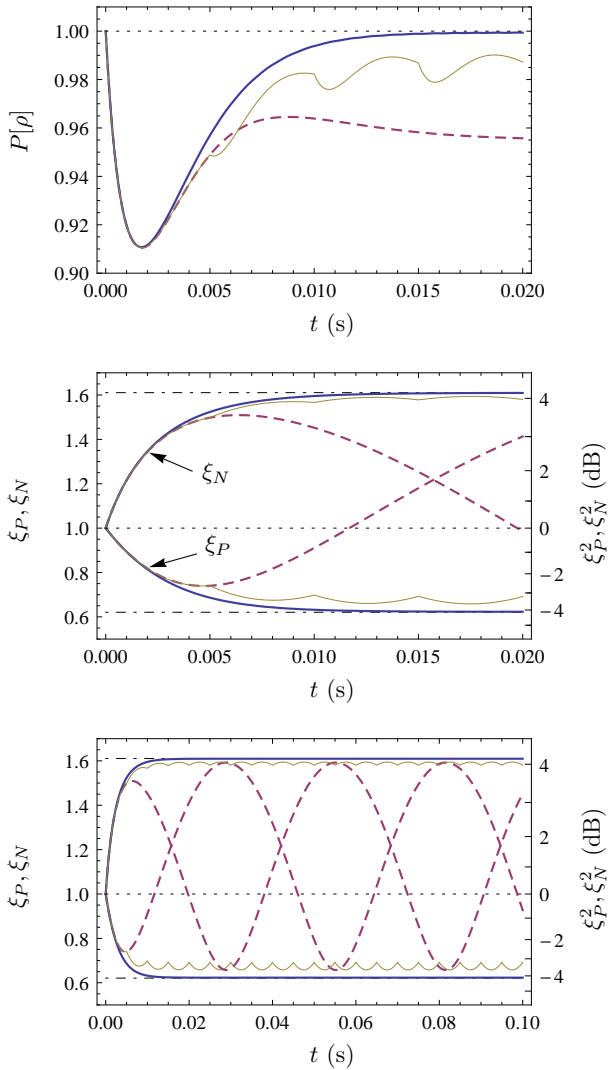}
\vspace{1mm}
\caption{(Color online) Purity and measures of phase and number squeezing for a 
system with $N=100$, $\gamma= 1.16\,\text{s}^{-1}$, $\omega=2\pi\times 19.0$ Hz, $\nu=-0.396$, and $\eta=1.12$. For these parameter values, $\tau_\gamma/T_\omega\approx 0.049$ and $\gamma_{\rm eff}\approx 391\,\text{s}^{-1}$, which are the same as those given in Table \ref{coeff}. The dashed line is obtained from Eq.~\eqref{masteq1}.
The thick solid line (thin solid line) is obtained using the stroboscopic method 
with $\tau_{\text{int}}=0.019 T_\omega=1$~ms ($\tau_{\text{int}}=0.095T_\omega=5$~ms). The horizontal dash-dotted lines are obtained using the analytical result (\ref{sssqueezingmeaszeroom}).
}
\label{fig:strobo}
\end{figure}

We start by presenting experimentally relevant values for the parameters of the system. 
They have been chosen to ensure that all the assumptions outlined in Sec.~II are satisfied. 
Our choice for the trapped atoms ($a$-atoms) 
is ${}^{85}$Rb. This makes it possible to use Feshbach resonances to 
set the scattering length of $a$-atoms to zero, which in turn 
helps to minimize the phase diffusion. 
We choose ${}^{87}$Rb as the background atoms ($b$-atoms). The values of the scattering length of the $b$-atoms $a_{bb}$ and the interspecies scattering length $a_{ab}$ are taken from Ref.~\cite{crimsonchin} and are given, together with the rest of the parameters, in Table \ref{sysparam}. The numerical values of the parameters appearing in the master equation are given in Table~\ref{coeff}. 
\begin{table}
\vspace{5mm}
\captionabove{Possible choice for the system parameters. Here $a$ refers to $^{85}$Rb, $b$ refers to $^{87}$Rb, and $a_0$ is the Bohr radius.  The values for $a_{bb}$ and $a_{ab}$ are obtained from Table IV of Ref.~\cite{crimsonchin}.}
\vspace{1mm}
               \begin{tabular}{>{\centering\arraybackslash}m{.24\textwidth}>{\centering\arraybackslash}m{.24\textwidth}}\hline\hline
          System parameter & Numerical value \\ \hline
          $m_{a}$ & $84.9\;\mathrm{a.u.}$\\
        $m_{b}$ & $86.9\;\mathrm{a.u.}$\\
      $a_{bb}$ & $100a_0$\\
        $a_{ab}$ & $213a_0$\\
            $\rho_{b}$ & $10^{21}\;\mathrm{m}^{-3}$\\
            $N$ & $10^5$\\
            $\sigma_g$ & $100\; \mathrm{nm}$\\
        $x_0$ & $350\; \mathrm{nm}$\\
        $\sigma_e$ & $2.5\; \mu\mathrm{m}$\\
      $\omega_{\rm well}$ & $2\pi\times11.9\; \mathrm{kHz}$\\
        $\omega=\epsilon_2-\epsilon_1$ & $2\pi\times19.0\; \mathrm{Hz}$\\
        $\epsilon_1$ & $2\pi\times 500\; \mathrm{Hz}$\\
             $|\Omega_1/\Delta_1|$ & $0.075$\\
        $|\Omega_2/\Delta_2|$ & $0.15$\\
         \hline\hline
        \end{tabular}
        \label{sysparam}
\end{table}
\begin{table}[t]
\vspace{5mm}
        \captionabove{Numerical values of the parameters appearing in the master equation \eqref{masteq1}, calculated using the values given in Table~\ref{sysparam}. Here $\eta$ is determined by the ratio $\epsilon_2/\epsilon_1$, $\gamma$ is the overall dissipation rate appearing in the master equation, $\nu$  controls  the amount of squeezing, and $\tau_{\gamma}$ and $T_\omega$ give the time scales characterizing the creation of squeezed states and the onset of dephasing, respectively.}
    \centering
    \vspace{1mm}
        \begin{tabular}{>{\centering\arraybackslash}m{.24\textwidth}>{\centering\arraybackslash}m{.24\textwidth}}\hline\hline
            Parameter & Numerical value \\ \hline
        $\eta$ & $1.12$\\
          $\gamma$& $0.00116\;\mathrm{s}^{-1}$\\
          $|\nu|$ & $0.396$\\
          $\gamma_{\rm eff}=\tau_{\gamma}^{-1}$ & $391\;\mathrm{s}^{-1}$ \\
          $\tau_{\gamma}/T_{\omega}$ & $0.049$\\ \hline\hline
        \end{tabular}
        \label{coeff}
\end{table}
The results of our numerical simulations are shown in Fig.~\ref{fig:strobo}.    
Since $\tau_\gamma/T_\omega\propto \omega/N\gamma$, increasing the number of particles makes the time interval  during which the system is in a squeezed state longer. 
Alternatively, the oscillatory behavior of $\xi_P$ and $\xi_N$ can be suppressed altogether by using a stroboscopic approach. In this approach, we let the state evolve via Eq.~(\ref{masteq1}) up to a short period of time $\tau_{\text{int}}\ll T_{\omega}$ so that we can avoid dephasing.
At $t=\tau_{\text{int}}$, we turn off the lasers for a very short interval of time, after which we turn the lasers on again, allowing the time evolution to continue for $\tau_{\text{int}}$. Repeating this procedure periodically, we can suppress the oscillations of $\xi_{P,N}$. 

\begin{center}
\end{center}

We show numerically obtained results for two values of $\tau_{\text{int}}$ in Fig.~\ref{fig:strobo}. In the numerical calculations, we set the length of the time interval during which the lasers are switched off to zero. We find that if $\tau_{\text{int}}$ is short enough, the dynamics is very close to that of a system with $\omega=0$. In particular, the steady-state value of squeezing can be almost identical in systems with $\omega=0$ and $\omega\neq 0$ if the stroboscopic method is used in the latter case. 
Replacing $d\hat{\rho}/dt$ with Eq.~(\ref{masteq1}) in the derivation of Eq.~(\ref{sssqueezingmeas}) and setting $\omega=0$, we find that the steady-state value reads 
\begin{equation}
\xi_{P,N}^{{\rm SS},\omega=0}\simeq\sqrt{\frac{1\pm\eta\nu}{1\mp\eta\nu}}.
\label{sssqueezingmeaszeroom}
\end{equation}
Here, the upper sign corresponds to the measure of phase squeezing and the lower sign to that of number squeezing. The thick solid line in Fig.~\ref{fig:strobo}, obtained using the stroboscopic method, can be seen to approach the horizontal dash-dotted lines corresponding to the steady-state values of the measure of squeezing given by Eq.~\eqref{sssqueezingmeaszeroom}. 
As $\tau_{\text{int}}$ becomes longer, the dynamics starts to deviate from that of a system with $\omega=0$ (see the thin solid lines in Fig.~\ref{fig:strobo}). The system remains in a squeezed state, but the amount of squeezing fluctuates in time  and the purity is smaller than one.

Even though the stroboscopic method makes it possible to effectively obtain $\omega=0$, the value of $\eta$ is not in general equal to one. This means that the dynamics deviates from that of the ideal case, characterized by $\omega=0$ and  $\eta=1$. 
From Eq.~\eqref{sssqueezingmeaszeroom} we see that the larger $\eta$ is, the more squeezing the stroboscopic method yields. 
 It should be noted, however, that an increase in $\eta$ leads to a decrease in purity unless $\tau_{\rm int}$ is short enough.
The main motivation for using the stroboscopic approach is that it allows us to obtain long-lived phase- and number-squeezed states using a relatively small number of particles.

\section{ Conclusions \label{sec:conclusion}}
We have worked out a dissipative scheme for  
the preparation of phase- and
number- squeezed states using trapped ultracold atoms in a double-well
setup immersed in a background BEC. Our scheme employs a coherent
superposition of two $\Lambda$ systems coupled to the background BEC. 
We derived a master equation for the system and found that the dynamics is accelerated by Bose statistics so that a squeezed state is achieved on a time scale 
$\sim 1/(N\gamma)$. We observed that this process 
suffers from dephasing on a time scale $\sim 1/\omega$ defined by the energy separation 
$\omega$ between the excited states corresponding to the
upper levels of the $\Lambda$ configurations. However, this dephasing
can be circumvented efficiently by combining the continuous time
evolution with a stroboscopic element. This leads to a robust
long-lived entanglement. 
Bosonic rubidium isotopes ${}^{85}$Rb and
${}^{87}$Rb offer favorable scattering properties for the
implementation of this scheme. In more detail, the trapped ${}^{85}$Rb atoms have a sufficiently
large interspecies scattering length with the ${}^{87}$Rb bath species and the scattering length 
between the ${}^{85}$Rb atoms confined in the double well can be set to zero using Feshbach resonances. 
In summary, we have shown that our dissipative preparation scheme for phase- and number-squeezed states using ultracold atoms 
is conceptually possible and we have 
provided the necessary ingredients for its experimental realization.

\begin{acknowledgments}
G.W. acknowledges the Max Planck Society; the Korea Ministry of
Education, Science and Technology; Gyeongsangbuk-Do; and Pohang City
for support of the JRG at APCTP. G.W. was also supported by Basic
Science Research Program through the National Research Foundation of
Korea (NRF) funded by the Ministry of Education, Science and
Technology (Grant No. 2012R1A1A2008028). R.C.F.C. acknowledges support from the University of the Philippines, the Centre for Quantum Technology, NITheP, and the University of KwaZulu-Natal. S.D. was supported  by the Austrian Science Fund (FWF) through the START Grant No. Y 581-N16 and the SFB FoQuS (FWF Project No. F4006-N16). 
H.M. was supported by the Alfred Kordelin Foundation, the Magnus Ehrnrooth Foundation, the Emil Aaltonen Foundation,
and the Academy of Finland through its Centres of Excellence Program (Project No. 251748).
\end{acknowledgments}
 
\appendix
\section{Field Operators and the Interaction Hamiltonian}\label{app:foh}
In deriving the interaction Hamiltonian given by Eq.~(\ref{inthamil}), we make use of the field operators for the trapped atoms $\hat{\psi}_a$ and the background BEC $\hat{\psi}_b$, given by

\begin{align}
  \hat{\psi}_a({\bf r}) =& \left[\phi_{g,1}(x) \hat{a}_{g,1} 
+ \phi_{g,2}(x) \hat{a}_{g,2}\right. \nonumber\\
&\left. + \phi_{e_1,0}(x) \hat{a}_{e_1,0}
+ \phi_{e_2,0}(x) \hat{a}_{e_2,0}
\right] w_y(y) w_z(z)
\end{align}
and
\begin{equation}
  \hat{\psi}_b = \sqrt{\rho_b} + \delta\hat{\psi}_b({\bf r})
\end{equation}
with
\begin{equation}
  \delta\hat{\psi}_b({\bf r}) = \frac{1}{\sqrt{V}} 
\sum_{{\bf k}} \left( u_{\bf k} \hat{b}_{\bf k} e^{i {\bf k}\cdot{\bf r}}
+ v_{\bf k} \hat{b}_{\bf k}^\dagger e^{-i {\bf k}\cdot{\bf r}}
\right) .
\end{equation}
Here, $w_y$ and $w_z$ are the ground state wave functions of the trap in the transverse directions, and the coefficients $u_{\bf k}$ and $v_{\bf k}$ are
$u_{\bf k}=1/\sqrt{1-L_{\bf k}^2}$ and $v_{\bf k}=L_{\bf k}/\sqrt{1-L_{\bf k}^2}$,
where $L_{\bf k}\equiv \left[E_{\bf k}-(k^2/2m_b)-m_bc^2\right]/m_bc^2$, 
$E_{\bf k} \equiv ck \sqrt{1+ (k/2m_bc)^2}$, 
$m_b$ is the mass of $b$-atoms, $c\equiv\sqrt{g_{bb}\rho_b/m_b}$ 
is the sound velocity in the background BEC, and $g_{bb}$ is the interaction
strength between $b$-atoms. Then $\hat{H}_{ab}$ is given by
\begin{equation}
  \hat{H}_{ab} = \frac{1}{2} \frac{4\pi a_{ab}}{\mu}
\int d^3r\ \hat{\psi}_a^\dagger\hat{\psi}_a \hat{\psi}_b^\dagger\hat{\psi}_b.
\end{equation}
In evaluating this integral, we consider only terms that are linear in $\hat{b}_{\bf k}$ and $\hat{b}_{\bf k}^{\dagger}$. This results in Eq.~(\ref{inthamil}).

\section{Evaluation of overlap integrals}\label{app:overlapint}
The overlap integrals can be simplified under some assumptions for the trap parameters. 
Let us first consider the integrals 
\begin{align}
\label{eq_overlap1}
&{\cal A}_{k_x;0,1}^{(e_1,g)}=\int dx\ e^{ik_x x} \phi_{e_1,0}^*(x)\phi_{g,1}(x)\\
\label{eq_overlap2}
&{{\cal A}_{k_x;0,2}^{(e_1,g)}}=\int dx\ e^{ik_x x} \phi_{e_1,0}^*(x)\phi_{g,2}(x).
\end{align}
Here $\phi_{e_1,0}$ is the ground state of a harmonic oscillator. It is therefore an 
even function, $\phi_{e_1,0}(x)=\phi_{e_1,0}(-x)$. Due to the symmetry of the 
setup the states $\phi_{g,1}$ and $\phi_{g,2}$ are related as $\phi_{g,1}(x)=\phi_{g,2}(-x)$. 
Using these equations and changing the integration variable as $x\to -x$ we get 
\begin{align}
\label{eq_overlap3}
&{\cal A}_{k_x;0,1}^{(e_1,g)}=\int dx\ e^{-i k_x x} \phi_{e_1,0}^*(x)\phi_{g,2}(x). 
\end{align}
Due to the $\delta$ functions appearing in Eq.~\eqref{master1}, the largest possible value of $k=\|{\bf k}\|$ is $k=k_2$ and consequently $-k_2\leq k_x\leq k_2$. The trap parameters are 
chosen in such a way that $k_2 x_0\ll 1$, where $x_0$ denotes the point around which $\phi_{g,2}$ 
is peaked. We also assume that $\phi_{g,2}$ is essentially zero outside the interval $(x_0-d,x_0+d)$, 
where $d\lesssim x_0$.
We then see that $ e^{-i k_x x}\simeq 1$ in the region where the integrand of the 
integral \eqref{eq_overlap3} is nonnegligible. Comparing Eqs.~\eqref{eq_overlap2} and \eqref{eq_overlap3}   
we see that 
\begin{align}
  {\cal A}_{k_1;0,1}^{(e_1,g)} &\simeq {{\cal A}_{k_1;0,2}^{(e_1,g)}}.
\end{align}
In a similar fashion, by noting that $\phi_{e_2,0}$ is an odd function, we get 
\begin{align}
{\cal A}_{k_2;0,1}^{(e_2,g)} &\simeq -{{\cal A}_{k_2;0,2}^{(e_2,g)}}. 
\end{align}
Having obtained the relationship between the overlap integrals, we now proceed to the evaluation of these integrals. Since $\sigma_g\ll\sigma_e$ and $\phi_{e_1,0}$ and $\phi_{e_2,0}$
are real functions, ${\cal A}_{k_x;0,i'}^{(e_n,g)} = \int dx\ e^{ik_xx} \phi_{e_n,0}^*(x)\phi_{g,i'}(x)\sim \sigma_g^{1/2} e^{ik_xx_0} \phi_{e_n,0}(x_0)$.
For concreteness, taking $\phi_{g, i}(x)$ as a Gaussian with width $\sigma_g$,
\begin{align}
\label{phig}
\phi_g^{\pm}(x)&=\frac{1}{\pi^{1/4}\sigma_g^{1/2}}
\exp{\left[-\frac{(x\pm x_0)^2}{2\sigma_g^2}\right]}\\
&\xrightarrow[\sigma_g\rightarrow 0]{} \sqrt{2}\pi^{1/4}\sigma_g^{1/2} \delta(x\pm x_0),
\end{align}
we obtain
\begin{align}
  {\cal A}_{k_x;0,i'}^{(e_n,g)} \simeq& \sqrt{2}\pi^{1/4}\sigma_g^{1/2} e^{\pm ik_xx_0}\phi_{e_n,0}(\pm x_0).
\end{align}
Here $+$ ($-$) corresponds to $i'=2$ $(i'=1)$.

\section{Integration over ${\bf k}$ and adiabatic elimination of $\hat{a}_{e_n}$}\label{app:integk}
The integration over ${\bf k}$ in Eq.~(\ref{masteq1}) will be performed using cylindrical coordinates in ${\bf k}$. 
Let $k^2 = k_\perp^2 + k_x^2$ with $k_\perp^2=k_y^2+k_z^2$. Integrating with respect to $k_\perp$ and the azimuthal angle of ${\bf k}$ in Eq.~(\ref{masteq1}) gives us
\begin{align}
  {\cal L}\hat{\rho} \simeq& \frac{1}{(2\pi)^3} \int d^3k\ |g_{\bf k}|^2
\left\{
[\hat{A}_{\bf k}(t),\hat{\rho}(t)\hat{A}_{\bf k}^\dagger(E_{\bf k},t)] + \mbox{h.c.}
\right\}\nonumber\\
  =& \sum_{n=1,2} \gamma_n \int_{-k_n}^{k_n} dk_x\ e^{-(k_n^2 - k_x^2)a_\perp^2 /2}
\nonumber\\
& \times\left[
e^{-i\epsilon_1 t}\left(
{{\cal A}_{k_x;0,1}^{(e_1,g)}}^* \hat{a}_{g,1}^\dagger 
+ {{\cal A}_{k_x;0,2}^{(e_1,g)}}^* \hat{a}_{g,2}^\dagger
\right) \hat{a}_{e_1,0} \right.\nonumber\\
& + e^{-i\epsilon_2 t}\left(
{{\cal A}_{k_x;0,1}^{(e_2,g)}}^* \hat{a}_{g,1}^\dagger 
+ {{\cal A}_{k_x;0,2}^{(e_2,g)}}^* \hat{a}_{g,2}^\dagger
\right) \hat{a}_{e_2,0},\nonumber\\
& \left.
\hat{\rho}e^{i\epsilon_n t}\hat{a}_{e_n,0}^\dagger \left({\cal A}_{k_x;0,1}^{(e_n,g)} \hat{a}_{g,1}
+ {\cal A}_{k_x;0,2}^{(e_n,g)} \hat{a}_{g,2} \right)
\right] + \mbox{h.c.}\label{eq_liuville}
\end{align}
\indent Then we perform the $k_x$ integration in Eq.\ (\ref{eq_liuville}). 
The $k_x$ dependence of the integrand in Eq.\ (\ref{eq_liuville})
comes from $e^{\pm ik_xx_0}$ in ${\cal A}_{k_x;i,i'}^{(n,n')}$ and the overall factor $e^{-(k_n^2 - k_x^2)a_\perp^2 /2}$. 
 Using the assumptions made about the system in Sec.~\ref{sec:sys}, we get $k_na_\perp\ll k_n x_0 \sim x_0/\sigma_e \ll 1$. 
This means that $e^{\pm ik_xx_0}\simeq 1$ and $e^{-(k_n^2 - k_x^2)a_\perp^2 /2}\simeq 1$ throughout the integration range and the $k_x$ integration simply yields $\int^{k_n}_{-k_n} dk_x \rightarrow 2k_n$.
Performing the $k_x$ integration, we obtain the following explicit form of the master equation:
\begin{align}
  {\cal L}\hat{\rho} =& 4\sqrt{\pi}k_1 \gamma_1 \sigma_g \phi_{e_1,0}^2(x_0)
\nonumber\\
&\times \biggl\{[\hat{c}_+^\dagger \hat{a}_{e_1,0},\, \hat{\rho} \hat{a}_{e_1,0}^\dagger \hat{c}_+]+\tilde{\nu}e^{-it(\epsilon_2 - \epsilon_1)}[\hat{c}_-^\dagger \hat{a}_{e_2,0},\, \hat{\rho} \hat{a}_{e_1,0}^\dagger \hat{c}_+]\nonumber\\
& \left. + \frac{k_2\gamma_2}{k_1\gamma_1}\left(
\tilde{\nu}^2[\hat{c}_-^\dagger \hat{a}_{e_2,0},\, \hat{\rho} \hat{a}_{e_2,0}^\dagger \hat{c}_-
]\right.\right.\nonumber\\
&\left.\left.+\tilde{\nu}e^{it(\epsilon_2-\epsilon_1)}[\hat{c}_+^\dagger \hat{a}_{e_1,0},\, \hat{\rho} \hat{a}_{e_2,0}^\dagger \hat{c}_-]\right)
\right\} + \mbox{h.c.}\label{masteqinter}
\end{align}
Next we perform the adiabatic elimination of $\hat{a}_{e_1,0}$ and $\hat{a}_{e_2,0}$. 
This can be done as the Raman lasers coupling the states of the double well ($\varphi_{g,1}$ and $\varphi_{g,2}$) 
to the states of the 
harmonic trap ($\varphi_{e_1,0}$ and $\varphi_{e_2,0}$) are weak and far detuned. After the  adiabatic  elimination of $\hat{a}_{e_1,0}$ and $\hat{a}_{e_2,0}$ we get 
\begin{align}
  \hat{a}_{e_1,0}\simeq& \frac{\Omega_1}{\sqrt{2}\Delta_1} 
(\hat{a}_{g,1}-\hat{a}_{g,2}),\label{adia1}\\
\hat{a}_{e_2,0}\simeq& \frac{\Omega_2}{\sqrt{2}\Delta_2} 
(\hat{a}_{g,1}+\hat{a}_{g,2}).\label{adia2}
\end{align}
Finally, by substituting Eqs.~(\ref{adia1}) and (\ref{adia2}) into Eq.~(\ref{masteqinter}) and simplifying the resulting equation using the notation introduced in Eqs.~(\ref{etaconst})--(\ref{cop}), we obtain the master equation given by Eq.~(\ref{masteq0}).


\begin{thebibliography}{99}
%
\bibitem{esteve} J. Est\`eve, C. Gross, A. Weller, S. Giovanazzi, and M.~K. Oberthaler, Nature (London) {\bf 455}, 1216-1219 (2008).
%
\bibitem{appel} J. Appel, P.~J. Windpassinger, D. Oblak, U.~B. Hoff, N. Kj{\ae}rgaard, and E.~S. Polzik, Proc.\ Nat.\ Acad.\ of Sci. USA {\bf 106}, 10960 (2009).
%
\bibitem{cronin} A. Cronin, J. Schmiedmayer, and D.~E. Pritchard, Rev.\ Mod.\ Phys. {\bf 81}, 1051 (2009).
%
\bibitem{leroux} I.~D. Leroux, M.~H. Schleier-Smith, and V. Vuleti\'c, Phys.\ Rev.\ Lett. {\bf 104}, 073602 (2010).
%
\bibitem{gross} C. Gross, T. Zibold, E. Nicklas, J. Est\`eve, and M.~K. Oberthaler, Nature (London) {\bf 464}, 1165 (2010).
%
\bibitem{riedel} M.~F. Riedel, P. B\"ohi, Y. Li, T.~W. H\"ansch, A. Sinatra, and P. Treutlein,
  Nature (London) {\bf 464}, 1170 (2010).
%
\bibitem{grond} J. Grond, U. Hohenester, I. Mazets, and J. Schmiedmayer, New J.\ Phys. {\bf 12}, 065036 (2010).
%
\bibitem{lee} C. Lee, J. Huang, H. Deng, H. Dai, and J. Xu, Front.\ Phys. {\bf 7}, 109 (2012).
%
\bibitem{poyatos}
J.~F. Poyatos, J.~I. Cirac, and P. Zoller, Phys.\ Rev.\ Lett. {\bf 77}, 4728 (1996).
%
\bibitem{daley04} A.~J. Daley, P.~O. Fedichev, and P. Zoller,
  Phys.\ Rev.\ A {\bf 69}, 022306 (2004).
%
\bibitem{diehl} S. Diehl, A. Micheli, A. Kantian, B. Kraus, H.~P. B\"uchler, and P. Zoller, Nat.\ Phys. {\bf 4}, 878 (2008)
%
\bibitem{verstraete} F. Verstraete, M.~M. Wolf, and J.~I. Cirac, Nat.\ Phys. {\bf 5}, 633 (2009).
%
\bibitem{allothers} A. Eckardt, C. Weiss, and M. Holthaus, Phys.\ Rev.\ Lett. {\bf 95}, 260404 (2005); C.~E. Creffield, {\it ibid.} {\bf 99}, 110501 (2007); F. Piazza, L. Pezz\'e, and A. Smerzi, Phys.\ Rev.\ A {\bf 78}, 051601(R) (2008); G. Watanabe, {\it ibid.} {\bf 81}, 021604 (2010); C. Ottaviani, V. Ahufinger, R. Corbal\'an, and J. Mompart, {\it ibid.} 81, 043621 (2010); M.~S. Rudner, L.~M.~K. Vandersypen, V. Vuleti\'c, and L.~S. Levitov, Phys.\ Rev.\ Lett. {\bf 107}, 206806 (2011); M.~A. Leung, K.~W. Mahmud, and W.~P. Reinhardt, Mol.\ Phys. {\bf 110}, 801 (2012); G. Watanabe and H. M\"akel\"a, Phys.\ Rev.\ A {\bf 85}, 053624 (2012).
%
\bibitem{krauter} H. Krauter, C.~A. Muschik, K. Jensen, W. Wasilewski, J.~M. Petersen, J.~I. Cirac, and E.~S. Polzik, Phys.\ Rev.\ Lett. {\bf 107}, 080503 (2011).
%
\bibitem{diehl2} S. Diehl, A. Tomadin, A. Micheli, R. Fazio, and P. Zoller, 
  Phys.\ Rev.\ Lett. {\bf 105}, 015702 (2010).
%
\bibitem{weimer} H. Weimer, M. M\"uller, I. Lesanovsky, P. Zoller, and H. P. B\"uchler, Nat.\ Phys. {\bf 6}, 382 (2010).
%
\bibitem{diehlt}
S. Diehl, E. Rico, M.~A. Baranov, and P. Zoller, Nat.\ Phys. {\bf 7}, 971 (2011).
%
\bibitem{pastawski}
F. Pastawski, L. Clemente, and J.~I. Cirac, Phys.\ Rev.\ A {\bf 83}, 012304 (2011).
%
\bibitem{kastoryano}
M.~J. Kastoryano, M.~M. Wolf, and J. Eisert, Phys.\ Rev.\ Lett. {\bf 110}, 110501 (2013).
%
\bibitem{kasto} M.~J. Kastoryano, F. Reiter, and A.~S. S{\o}rensen, Phys.\ Rev.\ Lett. {\bf 106}, 090502 (2011).
%
\bibitem{chen} X.~Y. Chen, L.~T. Shen, Z.~B. Yang, H.~Z. Wu, and M.~F. Chen, J.\ Opt.\ Soc.\ Am. B {\bf 29}, 1535 (2012).
%
\bibitem{marcos}  D. Marcos, A. Tomadin, S. Diehl, and P. Rabl, New J.\ Phys. {\bf 14}, 055005 (2012).
%
\bibitem{reiter} F. Reiter, M.~J. Kastoryano, and A.~S. S{\o}rensen, New J.\ Phys. {\bf 14}, 053022 (2012).
%
\bibitem{squeeze} G. Watanabe and H. M\"akel\"a, Phys.\ Rev.\ A {\bf 85}, 023604 (2012).
%
\bibitem{lukin} E.~G. Dalla Torre, J. Otterbach, E. Demler, V. Vuletic, and M.~D. Lukin,  Phys.\ Rev.\ Lett. {\bf 110}, 120402 (2013).
%
\bibitem{wstate} R. Sweke, I. Sinayskiy, and F. Petruccione, Phys.\ Rev.\ A {\bf 87}, 042323 (2013).
%
\bibitem{muschikth}
C.~A. Muschik, E.~S. Polzik, and J.~I. Cirac, Phys.\ Rev.\ A {\bf 83}, 052312 (2011).
%
\bibitem{muschik}
C.~A. Muschik, H. Krauter, K. Jensen, J.~M. Petersen, J.~I. Cirac, and E.~S. Polzik, J.\ Phys.\ B: At.\ Mol.\ Opt.\ Phys. {\bf 45}, 124021 (2012).
%
\bibitem{barreiroexp}
J.~T. Barreiro, M. M\"uller, P. Schindler, D. Nigg, T. Monz, M. Chwalla, M. Hennrich, C.~F. Roos, P. Zoller, and R. Blatt, Nature (London) {\bf 470}, 486 (2011).
%
\bibitem{schindlerexp}
P. Schindler, M. M\"uller, D. Nigg, J.~T. Barreiro, E.~A. Martinez, M. Hennrich, T. Monz, S. Diehl, P. Zoller, and R. Blatt,
Nat.\ Phys. {\bf 9}, 361 (2013).
%
\bibitem{note:atomspecies}
The trapped atoms and background BEC atoms need to experience a different trapping potential. 
Furthermore, we set the scattering length of the trapped atoms to zero using Feshbach resonances.  The scattering length of the background BEC atoms, in contrast, is taken to be positive. 
For these reasons, we assume that the trapped atoms and background BEC atoms are of different species. 
%
\bibitem{note:noham}
In the double-well setting described by the two-site Bose-Hubbard
Hamiltonian, for example, this can be achieved by making the potential
barrier between the wells high enough so that the tunneling is
suppressed and by using Feshbach resonances to reduce the scattering
length so that the on-site interaction strength can be zero.
%
\bibitem{note:epsilon} 
The parameter $\nu$ in the region of $|\nu|<1$ yields the squeezing around the coherent state with relative phase 0. In contrast, $\nu$ in the region of $|\nu|>1$ yields the squeezing around the coherent state with relative phase $\pi$. In this case, the steady-state values of $\xi_P$ and $\xi_N$ are given by $\xi_P^{\rm SS}=\sqrt{(1+\nu^{-1})/(1-\nu^{-1})}$ and $\xi_N^{\rm SS}=\sqrt{(1-\nu^{-1})/(1+\nu^{-1})}$, respectively [cf. Eq.~\eqref{sssqueezingmeas}] and the time constant $\tau_\gamma$ of the equations of motion of $\langle \hat{S}_{y,z}^2\rangle$ is given by $\tau_\gamma = 1/[4N\gamma(\nu^2-1)]$ [cf. Eq.~\eqref{chartime}].
In the present paper, however, we mainly consider the case of $|\nu|<1$.
%
\bibitem{note:temperature}
In our case (see Sec.~\ref{sec:results}), $\epsilon_1 = 2\pi\times 500$ Hz, corresponding to $\simeq 24$ nK. Temperatures lower than this value have been achieved and detected in experiments [see, for example, R. Gati, B. Hemmerling, J. F\"olling, M. Albiez, and M.~K. Oberthaler, Phys. Rev. Lett. \textbf{96}, 130404 (2006)].
%
\bibitem{note:overlap}
Under the two-mode approximation and in the double-well setting considered here, the tunneling Hamiltonian reads $\hat{H}_t=-J(\hat{a}_1^\dag\hat{a}_2+\hat{a}_2^\dag\hat{a}_1)$ with $J\equiv \int dx\ \phi_g^{\pm}(x) \left[(-\hbar^2/2m_a)\nabla^2+V_{\rm trap}(x)\right] \phi_g^{\mp}(x)$, where $\phi_g^{\pm}(x)$ are defined in Eq.~\eqref{phig} and $V_{\rm trap}(x)$ is the trap potential shown in Fig.~\ref{fig_setup}, which is well approximated by a harmonic potential with frequency $\omega_{\rm well}$ for each well. Thus, using the parameter values given in Table~\ref{sysparam}, $J$ is estimated as $J\sim \hbar\omega_{\rm well} e^{-x_0^2/\sigma_g^2} = O(10^{-1})$ s$^{-1}$. 
This is much smaller than the effective rate of squeezing $\gamma_{\rm eff}=391\, \mathrm{s}^{-1}$. Hence tunneling is negligible on the time scale $\tau_\gamma=\gamma_{\rm eff}^{-1}$ over which squeezing builds up. 
%
\bibitem{note:epsilon1}
Note that $\tau_\gamma$ diverges at $|\nu|=1$. Physically this means that we cannot achieve either $\xi_P=0$ or $\xi_N=0$ with the squeezing jump operator if the particle number $N$ is finite.
%
\bibitem{crimsonchin} C. Chin, R. Grimm, P. Julienne, and E. Tiesinga, Rev.\ Mod.\ Phys. {\bf 82}, 1225 (2010).
\end{thebibliography}
\end{document}